\documentclass[letterpaper]{article} 
\usepackage{aaai2026}  
\usepackage{times}  
\usepackage{helvet}  
\usepackage{courier}  
\usepackage[hyphens]{url}  
\usepackage{graphicx} 
\usepackage{natbib}  
\usepackage{caption} 
\usepackage{algorithm}
\usepackage{algorithmic}
\usepackage{mathtools}
\usepackage{amsfonts}
\usepackage{xcolor}

\usepackage{booktabs}
\usepackage{array}
\usepackage{multirow}
\usepackage{enumitem}

\usepackage{newfloat}
\usepackage{listings}
\DeclareCaptionStyle{ruled}{labelfont=normalfont,labelsep=colon,strut=off} 
\floatstyle{ruled}
\newfloat{listing}{tb}{lst}{}
\floatname{listing}{Listing}
\nocopyright

\title{CoCoLIT: ControlNet-Conditioned Latent Image Translation\\for MRI to Amyloid PET Synthesis}
\author {
    Alec Sargood\textsuperscript{* \rm 1},
    Lemuel Puglisi\textsuperscript{* \rm 2},\\
    James H. Cole\textsuperscript{\rm 1},
    Neil P. Oxtoby\textsuperscript{\rm 1},
    Daniele Ravì\textsuperscript{$\dagger$ \rm 3},
    Daniel C. Alexander\textsuperscript{$\dagger$ \rm 1}\\
}
\affiliations {
    \textsuperscript{*} Joint First Authors, $\dagger$ Joint Senior Authors
    
    \textsuperscript{\rm 1}Hawkes Institute, University College London, UK\\
    \textsuperscript{\rm 2}Dept. of Math and Computer Science, University of Catania, Italy\\
    \textsuperscript{\rm 3}MIFT Department, University of Messina, Italy\\
    alec.sargood.23@ucl.ac.uk, lemuel.puglisi@phd.unict.it\\
}

\usepackage{bibentry}

\begin{document}

\maketitle

\begin{abstract}
Synthesizing amyloid PET scans from the more widely available and accessible structural MRI modality offers a promising, cost-effective approach for large-scale Alzheimer's Disease (AD) screening. This is motivated by evidence that, while MRI does not directly detect amyloid pathology, it may nonetheless encode information correlated with amyloid deposition that can be uncovered through advanced modeling. However, the high dimensionality and structural complexity of 3D neuroimaging data pose significant challenges for existing MRI-to-PET translation methods. Modeling the cross-modality relationship in a lower-dimensional latent space can simplify the learning task and enable more effective translation. As such, we present CoCoLIT (ControlNet-Conditioned Latent Image Translation), a diffusion-based latent generative framework that incorporates three main innovations: (1) a novel Weighted Image Space Loss (WISL) that improves latent representation learning and synthesis quality; (2) a theoretical and empirical analysis of Latent Average Stabilization (LAS), an existing technique used in similar generative models to enhance inference consistency; and (3) the introduction of ControlNet-based conditioning for MRI-to-PET translation. We evaluate CoCoLIT's performance on publicly available datasets and find that our model significantly outperforms state-of-the-art methods on both image-based and amyloid-related metrics. Notably, in amyloid-positivity classification, CoCoLIT outperforms the second-best method with improvements of +10.5\% on the internal dataset and +23.7\% on the external dataset.
\end{abstract} 

\begin{links}
    \link{Code}{https://github.com/brAIn-science/CoCoLIT}
\end{links}


\section{Introduction}
Alzheimer's Disease (AD) places a substantial burden on patients, their families, and healthcare systems globally. As the population continues to age, both the human and economic costs associated with AD are steadily increasing~\cite{tay2024economic}. Early and accurate diagnosis is critical for effective intervention in AD. Among the available neuroimaging techniques, amyloid (A$\beta$) Positron Emission Tomography (PET) plays a key role by detecting A$\beta$ plaque accumulation—an early hallmark of AD—often years before cognitive symptoms appear~\cite{nordberg2004pet}. This makes A$\beta$ PET a vital tool for both research and clinical applications where early and reliable diagnosis is essential. However, the high cost and limited availability of A$\beta$ PET~\cite{lee2021cost}, as well as the radiation exposure, hinder its widespread use as a routine diagnostic tool. In contrast, structural Magnetic Resonance Imaging (MRI) is a more affordable, non-invasive, and widely used modality; however, it is less effective for early AD diagnosis, as it is not designed to highlight A$\beta$ plaques. Despite this limitation, MRI can still capture hidden information related to A$\beta$ pathology~\cite{kerbler2015basal}. While not a direct replacement for the biochemical accuracy of PET imaging, synthesizing A$\beta$ PET scans from structural MRI is a promising method for enabling large-scale, cost-effective AD screening, especially in resource-limited or low-income countries \cite{chapleau2022role}.

Extensive research has explored translating structural MRI data into PET images, with many approaches leveraging Generative Adversarial Networks (GANs) for their ability to synthesize realistic outputs. Notably, in~\cite{pan2018synthesizing}, the authors propose 3D-cGAN by extending the CycleGAN architecture~\cite{zhu2017unpaired} to 3D, enabling unpaired PET synthesis from MRI. Similarly, in~\cite{shin2020gandalf}, the authors build on the well-known pix2pix framework~\cite{isola2017image} for paired MRI-to-PET translation. While GAN-based methods have shown promising results in generating visually plausible images, they remain prone to training instabilities and mode collapse. Denoising Diffusion Probabilistic Models (DDPMs)~\cite{ho2020denoising} have recently emerged as powerful generative models capable of synthesizing high-fidelity, diverse images through a learned denoising process. Consequently, they have been adopted in State-of-the-Art (SOTA) methods for MRI-to-PET synthesis. For instance, FICD~\cite{yu2024functional} employs a conditional diffusion model that integrates an additional imaging constraint during training to enhance the fidelity and clinical relevance of the generated PET images. However, performing the diffusion process in the 3D image space imposes significant computational demands. Another recent method, PASTA~\cite{li2024pasta}, introduces a pathology-aware conditional diffusion model with an additional cycle exchange consistency loss. While PASTA addresses the challenges of 3D data by operating on sets of 2D slices, this limits the model's ability to fully capture inter-slice dependencies. To mitigate the challenges of image-space modeling, the authors of~\cite{ou2024image} propose IL-CLDM, a diffusion-based MRI-to-PET translation model that operates in a learned latent space. During training, the model is conditioned on the A$\beta$-positivity label by adding a learned label embedding to the time-step embedding. However, at inference time, the label is unavailable, and only the time-step embedding is used. This mismatch between training and inference conditions may lead to out-of-distribution behavior during generation.

\subsection{Contributions}
To address the limitations of prior work, we present CoCoLIT (\textbf{Co}ntrolNet-\textbf{Co}nditioned \textbf{L}atent \textbf{I}mage \textbf{T}ranslation), a diffusion-based model for conditional 3D medical image synthesis, focused on MRI-to-PET translation. CoCoLIT builds on recent advances in generative modeling, including latent diffusion~\cite{rombach2022high} and ControlNets~\cite{zhang2023adding}. Our key contributions are threefold: \textbf{(1)} we introduce a novel Weighted Image Space Loss (WISL), which improves latent representation learning and enhances the fidelity of synthesized images; \textbf{(2)} we provide the first formal justification and empirical evaluation of Latent Average Stabilization (LAS)—originally proposed in~\cite{puglisi2025brain}—showing that while LAS is asymptotically biased, its bias becomes negligible in sufficiently well-trained models; and \textbf{(3)} we are the first to successfully apply a ControlNet-based model to the task of MRI-to-PET translation. Our results show that CoCoLIT achieves SOTA performance on both image-based and clinical metrics, significantly outperforming existing methods on internal and external test sets.


\section{Preliminaries}
In this section, we introduce the background on which CoCoLIT is built, including Latent Diffusion Models (LDMs), the ControlNet conditioning mechanism, and the LAS technique.

\subsection{Latent Diffusion Models}\label{sec:ldm} 
An LDM~\cite{rombach2022high} is a deep generative model used to learn a target data distribution in a compressed latent space, comprising a forward and reverse Markovian diffusion process. Input images, $x$, are first encoded into a latent representation $z$ using an encoder $\mathcal{E}$. Gaussian noise is incrementally added to the latent vector in the forward process over $T$ steps, starting from $z_0=z$. At each step $t$, noise is added to $z_{t-1}$ by sampling from the Gaussian transition probability $q(z_t|z_{t-1}) = \mathcal{N}(z_t; \sqrt{1 - \beta_t}\, z_{t-1}, \beta_t I)$, where $\beta_t$ follows a predefined variance schedule. This ensures that $z_T$ asymptotically approaches pure Gaussian noise. The reverse process aims to revert each diffusion step, allowing the generation of a latent embedding from the target distribution starting from pure noise $z_T$. The reverse transition probability has a Gaussian closed form, $q(z_{t-1}|z_t, z_0) = \mathcal{N}(z_{t-1}|\tilde\mu(z_0, z_t), \tilde\beta_t)$, conditioned on the ground-truth latent $z_0$. The mean $\tilde\mu(z_0, z_t)$ can be reparameterized in terms of $z_t$ and a noise term $\epsilon$. Therefore, a neural network, $\epsilon_\theta(z_t, t)$, is trained to predict the noise by optimizing the following objective \cite{ho2020denoising}:
\begin{equation}
\label{eqn:ddpmloss}
\mathcal{L}_{LDM} := \mathbb{E}_{t, z_t, \epsilon \sim \mathcal{N}(0, I)} \left[ 
\lVert \epsilon - \epsilon_\theta(z_t, t) \rVert^2 \right].
\end{equation}
For $\epsilon_\theta$, we employ a U-Net-based denoising network, which allows the use of a ControlNet mechanism~\cite{zhang2023adding} for conditional generation, as described in the next section.

\subsection{Conditioning ControlNet}
\label{sec:controlnet}
A ControlNet mechanism~\cite{zhang2023adding} enables a pre-trained diffusion model to be conditioned on an additional signal by injecting information into the intermediate layers of its U-Net. Specifically, each U-Net encoder layer $\mathcal{F}(\cdot\,; \Theta)$ is kept frozen, and a trainable copy $\mathcal{F}(\cdot\,; \Theta_c)$ is introduced to learn the conditioning signal. The outputs of this trainable copy are integrated back into the corresponding frozen block via zero-initialized convolutional layers, denoted as $\mathcal{Z}$. These zero-convolution layers ensure that at the start of training, the ControlNet does not alter the original model's behavior. As training progresses, the parameters of $\mathcal{Z}$ adapt to effectively inject the conditioning information. Formally, for a given layer with input $v$, frozen layer output $w = \mathcal{F}(v; \Theta)$, and conditioning signal $z_c$, the modified output $w_{CN}$ is computed as:
\begin{equation}
w_{CN} = w + \mathcal{Z}\big( \mathcal{F}(v + \mathcal{Z}(z_c; \Theta_{z1}); \Theta_c); \Theta_{z2} \big),
\end{equation}
The entire denoising network is denoted as $\epsilon_{\theta,\phi}(z_t, t; z_c)$, where $\theta$ are the frozen U-Net weights and $\phi = \{\Theta_c, \Theta_{z1}, \Theta_{z2}\}$ is the set of learnable ControlNet parameters. The ControlNet is trained by minimizing the standard diffusion loss:
\begin{equation}
\label{eqn:controlnet}
\mathcal{L}_{CN} := \mathbb{E}_{t, z_t, \epsilon \sim \mathcal{N}(0, I)} \left[
\lVert \epsilon - \epsilon_{\theta,\phi}(z_t, t; z_c) \rVert^2 \right].
\end{equation}

\subsection{Latent Average Stabilization}
\label{sec:las-prelim}
A latent conditional generative model aims to learn a conditional distribution $p(z^{(y)}|z^{(x)})$, where $z^{(x)}$ and $z^{(y)}$ are latent variables from paired input and output images $x$ and $y$, respectively. These latents are computed via encoders, $z^{(x)} = \mathcal{E}^{(x)}(x)$ and $z^{(y)} = \mathcal{E}^{(y)}(y)$, and are recovered back to image space via decoders $x=\mathcal{D}^{(x)}(z^{(x)})$ and $y=\mathcal{D}^{(y)}(z^{(y)})$. Generation is performed by sampling from the learned distribution $p(z^{(y)}|z^{(x)})$, and decoding the samples via $\mathcal{D}^{(y)}$. As such, the process of inferring $y$ is inherently stochastic, with randomness introduced by the sampling procedure. Therefore, for a given input $x$, we aim to compute the expectation over the decoded samples. This expectation is estimated using the sample mean, an unbiased estimator computed over $N$ samples:
\begin{equation*}
\bar{y} = \frac{1}{N}\sum_{j=1}^N \mathcal{D}^{(y)}\left(z^{(y,j)}\right),
\end{equation*}
where each $z^{(y,j)}$ is a sample from $p(z^{(y)}|z^{(x)})$. 
A practical drawback of this method is its computational cost, requiring $N$ forward passes through the decoder. To resolve this issue, in~\cite{puglisi2024enhancing,puglisi2025brain}, the authors propose LAS, which involves taking $m$ samples from the learned latent distribution, $z^{(y, 1)},\dots,z^{(y, m)}\sim p(z^{(y)}|z^{(x)})$, and decoding their sample mean, $\bar{z}^{(y)}$:
$$
\hat{y}=\mathcal{D}^{(y)}\left(\bar{z}^{(y)}\right), \quad \text{for} \quad \bar{z}^{(y)}=\frac{1}{m}\sum_{j=1}^m z^{(y, j)},
$$
requiring only one forward pass of the decoder. It is shown in~\cite{puglisi2025brain} that LAS, when applied to a spatiotemporal disease progression modeling task, substantially improves results across a wide range of metrics. As the authors in~\cite{puglisi2025brain} do not examine its statistical properties, we aim to fill this gap by providing a theoretical analysis of LAS and justifying its use as a reliable estimator in conditional generative tasks.


\begin{figure*}[t]
\centering
\includegraphics[width=\textwidth]{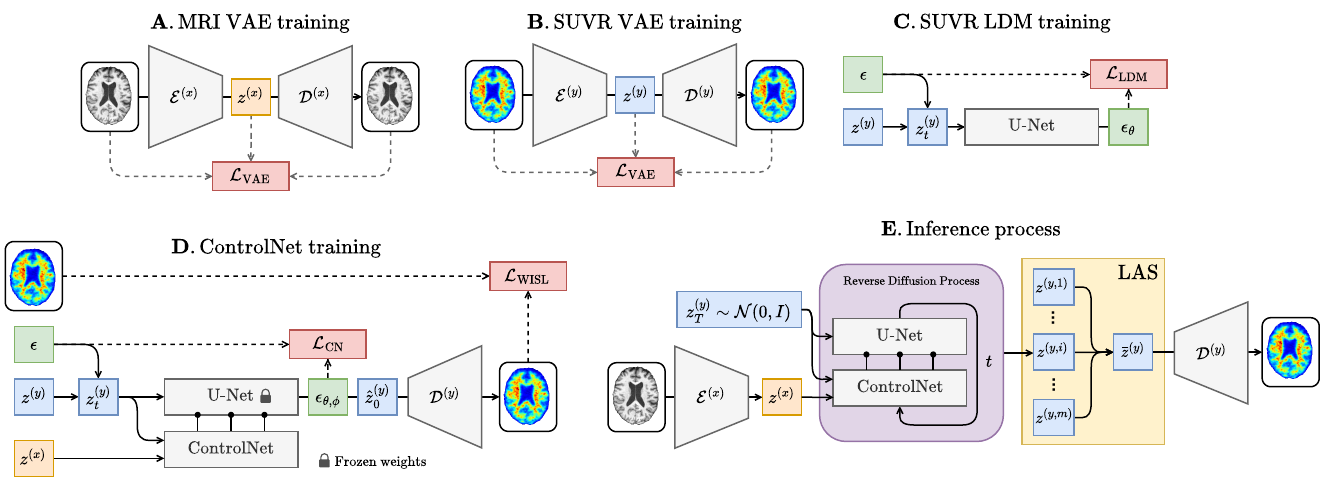}
    \caption{Overview of the CoCoLIT framework. (A–B) Training of the MRI and PET VAEs. (C) Training of the unconditional LDM on PET latents. (D) Training of the ControlNet and fine-tuning of the PET VAE decoder using standard noise loss and WISL. (E) Inference process in CoCoLIT, including the LAS algorithm.}
\label{fig:pipeline}
\end{figure*}

\section{Methods}
In this section, we describe the CoCoLIT framework, outlining the staged training process, our novel loss term (WISL), and a theoretical analysis of LAS.

\subsection{Proposed Framework: CoCoLIT}
\label{sec:proposed}
The overall pipeline of CoCoLIT, illustrated in Figure~\ref{fig:pipeline}, comprises five main blocks: (A–B) independent VAEs for MRI and PET representation learning, (C) an LDM for modeling latent PET distributions, (D) a ControlNet for conditional generation, and (E) the inference process incorporating LAS. Blocks A–D correspond to the training stages, while block E details inference. For further clarity, the inference process is also schematically described in Algorithm~\ref{alg:inference}.

\subsubsection{Representation Learning Stage}
This stage involves independently training two VAEs with the same architecture to encode and reconstruct 3D brain images. The MRI VAE (block A) encodes an input MRI volume $x\in\mathbb{R}^D$ into a latent representation $z^{(x)}\in\mathbb{R}^d$ using an encoder $\mathcal{E}^{(x)}$, and reconstructs the original image through a decoder $\mathcal{D}^{(x)}$. Similarly, the PET VAE (block B) maps an A$\beta$ PET scan $y\in\mathbb{R}^D$ into a latent code $z^{(y)}\in\mathbb{R}^d$ via encoder $\mathcal{E}^{(y)}$, and reconstructs it using decoder $\mathcal{D}^{(y)}$. Both VAEs are trained using a composite loss function, $\mathcal{L}_{\text{VAE}}$, which includes reconstruction, perceptual and adversarial losses, and a Kullback-Leibler regularization term, following the formulation in~\cite{guo2025maisi}. 

\subsubsection{Conditional Generative Modeling Stage}
The second stage starts by training an LDM (block C), which learns an unconditional distribution over the latents, $z^{(y)}$ (see Section~\ref{sec:ldm}). The final component (block D) is a ControlNet module, denoted as $\epsilon_{\theta,\phi}$, which operates on top of the trained and frozen LDM backbone (see Section~\ref{sec:controlnet}), thereby learning a conditional distribution, $p(z^{(y)}|z^{(x)})$. To improve image synthesis quality, we propose to incorporate image-space guidance by adding a loss term, which we call WISL, defined as:
\begin{equation}
\label{eqn:wist}
\mathcal{L}_{WISL}:=\mathbb{E}_{t, z_t^{(y)}, \epsilon \sim \mathcal{N}(0, I)} \left[\lambda_t\  \left\lVert y-\mathcal{D}^{(y)}(\hat{z}_0^{(y)}) \right\rVert_1 \right]. 
\end{equation}

Here, we calculate the weighted difference between the ground-truth PET $y$ and the decoded prediction $\mathcal{D}^{(y)}(\hat{z}_0^{(y)})$. The term $\hat{z}_0^{(y)}$ is an estimate of the fully denoised latent, recovered from the noised latent $z_t^{(y)}$ 
at time-step $t$~\cite{ho2020denoising}, and is given by the formula:
$$
\hat{z}_0^{(y)}=\left(z_t^{(y)}-\sqrt{1-\bar{\alpha}_t}\epsilon_{\theta,\phi}(z_t^{(y)}, t;  z^{(x)})\right)\cdot\left(\sqrt{\bar{\alpha}_t}\right)^{-1},
$$
where $\bar{\alpha}_t=\prod_{s=1}^t(1-\beta_s)$. 
We adopt a time-step-dependent weighting $\lambda_t \in [0,1]$ to scale the image-space loss, prioritizing low-frequency synthesis at high $t$ and high-frequency detail reconstruction at low $t$, in line with the progressive refinement process of the diffusion model~\cite{ho2020denoising}. For simplicity, we use a linear schedule defined as $\lambda_t = (T - t)/T$. The final loss term used in ControlNet training is given as:
\begin{equation}
\label{eqn:cn}
    \mathcal{L}_{WCN}=\mathcal{L}_{WISL} + \mathcal{L}_{CN}. 
\end{equation}
Since the loss term $\mathcal{L}_{WISL}$ is dependent on the decoder network, $\mathcal{D}^{(y)}$, we allow the decoder weights to be fine-tuned during ControlNet training (block D).

\begin{algorithm}[tb]
\caption{CoCoLIT Inference Procedure}
\label{alg:inference}
\textbf{Input}: MRI volume $x$, LAS hyperparameter $m$ \\
\textbf{Output}: Estimated PET scan $\hat{y}$
\begin{algorithmic}[1]
\STATE Encode MRI into latent space: $z^{(x)} = \mathcal{E}^{(x)}(x)$
\FOR{$j = 1$ to $m$}
    \STATE Sample Gaussian noise: $z_T^{(y, j)} \sim \mathcal{N}(0, I)$
    \FOR{$t = T$ to $1$}
        \STATE Reverse $z_t^{(y, j)} \to z_{t-1}^{(y, j)}$ using $\epsilon_{\theta,\phi}(z_t^{(y, j)}, t; z^{(x)})$
    \ENDFOR
    \STATE Store final latent: $z^{(y, j)} = z_0^{(y, j)}$
\ENDFOR
\STATE Compute $\bar{z}^{(y)} = \frac1m \sum_{j=1}^m z^{(y,j)}$
\STATE Decode PET scan: $\hat{y} = \mathcal{D}^{(y)}(\bar{z}^{(y)})$
\STATE \textbf{return} $\hat{y}$
\end{algorithmic}
\end{algorithm}

\subsection{Theoretical Analysis of LAS}
\label{sec:las}
In this section, we present an analysis of the LAS estimator, $\hat{y}$ (see Section~\ref{sec:las-prelim}), to characterize its bias and assess its statistical validity. A full derivation with additional details can be found in the Supplementary Material.

Let $\mu = \mathbb{E}[z^{(y)}|z^{(x)}]$ denote the conditional mean of the learned latent distribution, $p(z^{(y)}|z^{(x)})$. For notational simplicity, we denote $z^{(y)}$ as a sample from the conditional distribution $p(z^{(y)}|z^{(x)})$ throughout the rest of this section. We begin by establishing an approximation, via a second-order Taylor expansion of the decoder $\mathcal{D}^{(y)}$ around $\mu$, assuming that $p(z^{(y)}|z^{(x)})$ has finite second moments and $\mathcal{D}^{(y)}$ is twice continuously differentiable:
\begin{equation}
\label{taylor}
\begin{split}
\mathbb{E}\left[\mathcal{D}^{(y)}\left(z^{(y)}\right)\right] &\approx\ \mathbb{E}\left[\mathcal{D}^{(y)}(\mu)\right] \\
&+ \mathbb{E}\left[\nabla\mathcal{D}^{(y)}(\mu)(z^{(y)}-\mu)\right] \\
& + \frac{1}{2}\mathbb{E}\left[\left(z^{(y)}-\mu\right)^T H_{\mathcal{D}^{(y)}} \left(z^{(y)}-\mu\right)\right].
\end{split}
\end{equation}
Here, $H_{\mathcal{D}^{(y)}} \in \mathbb{R}^{D \times d \times d}$ denotes the Hessian tensor of the decoder evaluated at $\mu$, comprising one $d \times d$ Hessian matrix per output dimension. Since $\mathbb{E}[z^{(y)} - \mu] = 0$, the first-order term vanishes. Applying the linearity of expectation and the cyclic property of the trace operator yields:
\begin{equation}
\label{eqn:trace}
\mathbb{E}\left[\mathcal{D}^{(y)}\left(z^{(y)}\right)\right] \approx \mathcal{D}^{(y)}(\mu) + \frac{1}{2}\text{Tr}\left(H_{\mathcal{D}^{(y)}}\Sigma_{z^{(y)}}\right),
\end{equation}
where $\Sigma_{z^{(y)}} = \text{Cov}(z^{(y)}) = \mathbb{E}\left[(z^{(y)} - \mu)(z^{(y)} - \mu)^T\right]$ is the $d \times d$ covariance matrix. The term $\text{Tr}(H_{\mathcal{D}^{(y)}}\Sigma_{z^{(y)}})\in\mathbb{R}^D$ represents a vector, containing the trace of each $d\times d$ Hessian and covariance matrix multiplication. 

Applying the same approximation to the LAS estimator $\hat{y}$, and noting that for $m$ i.i.d. samples $\text{Cov}(\bar{z}^{(y)}) = \frac{1}{m}\Sigma_{z^{(y)}}$, its expectation is:
\begin{equation}
\label{eqn:estimator}
    \mathbb{E}\left[\hat{y}\right] = \mathbb{E}\left[\mathcal{D}^{(y)}\left(\bar{z}^{(y)}\right)\right] \approx \mathcal{D}^{(y)}(\mu) + \frac{1}{2m}\text{Tr}\left(H_{\mathcal{D}^{(y)}}\Sigma_{z^{(y)}}\right).
\end{equation}
The bias of the LAS estimator is therefore approximated by the difference between Eq. \eqref{eqn:estimator} and Eq. \eqref{eqn:trace}:
\begin{equation}
\label{eqn:bias}
    \text{Bias}(\hat{y}) \approx \left(\frac{1}{m}-1\right)\frac{1}{2}\text{Tr}\left(H_{\mathcal{D}^{(y)}}\Sigma_{z^{(y)}}\right).
\end{equation}
From Eq.~\eqref{eqn:bias}, we observe that as the number of latent samples $m \to \infty$, the bias does not vanish but instead converges to a constant:
$$ 
\lim_{m\to\infty} \text{Bias}(\hat{y}) = -\frac{1}{2}\text{Tr}\left(H_{\mathcal{D}^{(y)}}\Sigma_{z^{(y)}}\right). 
$$
This reveals that LAS is an asymptotically biased estimator of the expected output. However, we hypothesize that this bias is negligible in practice for a sufficiently well-trained latent generative model. The practical effectiveness of LAS is justified by the following core assumption about the model's properties:\newline

\noindent\textbf{Assumption 1}. \textit{The LAS estimator exhibits negligible bias under the assumption that the latent distribution induced by a well-trained conditional LDM is sufficiently concentrated, such that the decoder $\mathcal{D}^{(y)}$ is approximately linear within the support of the latent samples. This occurs when the covariance $\Sigma_{z^{(y)}}$ is small, restricting samples to a neighborhood where the decoder's curvature is negligible.}
\newline

If this condition holds, the asymptotic bias term will be close to zero. In Section~\ref{sec:las-linear}, we empirically show that the decoder behaves linearly within the sampled regions of latent space, and confirm LAS as an effective estimator. Therefore, despite its inherent bias, LAS can serve as a computationally efficient estimator for sufficiently well-trained models.


\begin{table*}[t!]
    \centering
    \begin{small}
    \begin{tabular}{@{}l rrr rr r@{}} 
    \toprule
    \multicolumn{1}{@{}l}{\textsc{\textbf{Setting}}} & \multicolumn{3}{c}{\textsc{\textbf{Image-based Metrics}}} & \multicolumn{3}{c}{\textsc{\textbf{A$\beta$-related Metrics}}} \\
    \cmidrule(lr){2-4} \cmidrule(lr){5-7}
    & \textsc{SSIM}$~\uparrow$ & \textsc{PSNR}$~\uparrow$ & \textsc{MSE}$~\downarrow$ & \textsc{CABC}$~\uparrow$ & \textsc{HABC}$~\uparrow$ & \textsc{BA}$~\uparrow$ \\
    \midrule
    \multicolumn{7}{@{}l}{\scshape\bfseries (A) Ablation on $m$} \\ \addlinespace
    $m=1$  & $0.865 \pm 0.047$  & $22.570 \pm 2.427$ & $0.0067 \pm 0.0051$ & $0.180~(p=0.006)$ & $0.334~(p<0.001)$ & $57.4$\% \\
    $m=2$  & $0.880 \pm 0.047$  & $23.175 \pm 2.575$ & $0.0059 \pm 0.0049$ & $0.210~(p=0.001)$ & $0.405~(p<0.001)$ & $52.1$\% \\
    $m=4$  & $0.889 \pm 0.049$  & $23.735 \pm 2.723$ & $0.0054 \pm 0.0048$ & $0.300~(p<0.001)$ & $0.369~(p<0.001)$ & $56.4$\% \\
    $m=8$  & $0.892 \pm 0.050$  & $23.936 \pm 2.738$ & \underline{$0.0051 \pm 0.0047$} & \underline{$0.306~(p<0.001)$} & $0.470~(p<0.001)$ & $57.1$\% \\
    $m=16$ & $0.894 \pm 0.050$  & $24.079 \pm 2.786$ & $\mathbf{0.0050 \pm 0.0047}$ & $0.292~(p<0.001)$ & $0.474~(p<0.001)$ & \underline{$60.6$\%} \\
    $m=32$ & \underline{$0.895 \pm 0.050$}  & \underline{$24.125 \pm 2.807$} & $\mathbf{0.0050 \pm 0.0047}$ & $0.287~(p<0.001)$ & \underline{$0.500~(p<0.001)$} & $56.7$\% \\
    $m=64$ & $\mathbf{0.896 \pm 0.050}$  & $\mathbf{24.135 \pm 2.820}$ & $\mathbf{0.0050 \pm 0.0047}$ & $\mathbf{0.328~(p<0.001)}$ & $\mathbf{0.522~(p<0.001)}$ & $\mathbf{62.3}$\% \\
    \midrule
    \multicolumn{7}{@{}l}{\scshape\bfseries (B) Component Ablation} \\ \addlinespace
    Base        & $0.841 \pm 0.054$ & $21.251 \pm 2.370$ & $0.0088 \pm 0.0053$ & $0.026~(p=0.694)$ & $0.253~(p<0.001)$ & $43.9$\% \\
    + ISL  & $0.870 \pm 0.050$ & $22.446 \pm 2.767$ & $0.0072 \pm 0.0057$ & $0.048~(p=0.476)$ & $0.280~(p<0.001)$ & \underline{$58.5$\%} \\
    + WISL & $0.865 \pm 0.047$ & $22.570 \pm 2.427$ & $0.0067 \pm 0.0051$ & $0.180~(p=0.006)$ & $0.334~(p<0.001)$ & $57.4$\% \\
    + LAS  & \underline{$0.887 \pm 0.040$} & $22.520 \pm 2.462$ & \underline{$0.0066 \pm 0.0038$} & $0.175~(p=0.008)$ & $\mathbf{0.545~(p<0.001)}$ & $56.4$\% \\
    + LAS + ISL & $\mathbf{0.896 \pm 0.051}$ & \underline{$24.030 \pm 2.681$} & $\mathbf{0.0050 \pm 0.0043}$ & \underline{$0.281~(p<0.001)$} & $0.422~(p<0.001)$ & $56.7$\% \\    
    + LAS + WISL & $\mathbf{0.896 \pm 0.050}$ & $\mathbf{24.135 \pm 2.820}$ & $\mathbf{0.0050 \pm 0.0047}$ & $\mathbf{0.328~(p<0.001)}$ & \underline{$0.522~(p<0.001)$} & $\mathbf{62.3}$\% \\

    \midrule
    \multicolumn{7}{@{}l}{\scshape\bfseries (C) Image vs. Latent Space Averaging} \\ \addlinespace
    Unb. Estm. ($\bar y$) & $0.896 \pm 0.050$ & $24.173 \pm 2.803$ & $0.0049 \pm 0.0047$ & $0.327~(p<0.001)$ & $0.541~(p<0.001)$ & $60.1$\% \\
    LAS ($\hat y$) & $0.896 \pm 0.050$ & $24.135 \pm 2.820$ & $0.0050 \pm 0.0047$ & $0.328~(p<0.001)$ & $0.522~(p<0.001)$ & $62.3$\% \\

    \bottomrule
    \end{tabular}
    \end{small}
    \caption{(A) Impact of varying LAS hyperparameter $m$ for CoCoLIT on the internal test set. (B) Contribution of different model components (ISL, WISL, LAS) evaluated on the internal test set. (C) Comparison of LAS with the unbiased estimator (Unb. Estm.). Image-based metrics (SSIM, PSNR, MSE) are reported as \textit{mean~$\pm$~std}. A$\beta$ metrics (CABC, HABC, BA) evaluate A$\beta$-burden correlation and classification performance, with p-values for CABC and HABC reported in parentheses. The best result is highlighted in bold, and the second-best is underlined.}
    \label{tab:ablation-study}
\end{table*}

\section{Experiments}
This section presents an evaluation of the proposed CoCoLIT framework. We begin by briefly describing the internal and external datasets used in this study, along with the evaluation protocol adopted. We then conduct an ablation study to determine the impact of the LAS hyperparameter $m$, as well as to quantify the contribution of each component within the CoCoLIT framework. Furthermore, we benchmark CoCoLIT against the SOTA methods in MRI-to-PET translation. Finally, we empirically assess the validity of the theory underpinning LAS.

\subsection{Datasets and Pre-processing}
For training and evaluation of our framework, we use two publicly available multimodal neuroimaging datasets: ADNI~\cite{petersen2010alzheimer} and the A4 Study (including the LEARN substudy)~\cite{sperling2014a4}. Both datasets contain paired T1-weighted MRI and Florbetapir PET scans. The ADNI dataset includes 1,515 paired scans from 787 subjects (mean age: $74.9 \pm 7.6$ years; 50.6\% female; 85.1\% A$\beta$-positive). We split this dataset into training (80\%), validation (5\%), and test (15\%) sets, ensuring strict subject-level separation to prevent data leakage. To assess generalization, we use the A4 cohort as an external test set, drawing a random sample of 350 image pairs from 350 unique subjects (mean age: $63.9 \pm 22.8$ years; 59.7\% female; 83.1\% A$\beta$-positive). Following standard practice~\cite{schreiber2015comparison,royse2021validation}, we convert PET scans to Standardized Uptake Value Ratio (SUVR) maps using the cerebellar gray matter as the reference region. Ground-truth $A\beta$-positivity is defined as the mean SUVR value in the cerebral cortex exceeding the commonly used threshold of 1.11~\cite{schreiber2015comparison,royse2021validation}. To retain subject-specific patterns and preserve inter-modality relationships, we perform a z-score standardization on both MRI and PET scans independently using statistics computed from the training set. We resample all MRI and SUVR volumes to a uniform spatial resolution of $1.5$ mm$^3$. Full pre-processing details are provided in the Supplementary Material.

\subsection{Implementation Details}\label{sec:implementation}
The MRI and PET VAEs used in Section~\ref{sec:proposed} were obtained by fine-tuning separate instances of the MAISI VAE~\cite{guo2025maisi}. This network was chosen for its extensive pre-training on large amounts of 3D medical imaging data. All the blocks in CoCoLIT were implemented using the MONAI framework~\cite{pinaya2023generative}, and all training and experiments were conducted on an NVIDIA A100 GPU. At inference time, latent samples were generated from the LDM using an implicit sampling strategy (DDIM)~\cite{song2020denoising}, with 50 inference steps. The full implementation of CoCoLIT is available at \url{https://github.com/brAIn-science/CoCoLIT}.

\subsection{Evaluation Protocol}
\label{sec:evaluation-metrics}
We assess model performance using six metrics: three image-based measures—Structural Similarity Index Measure (SSIM), Peak Signal-to-Noise Ratio (PSNR), and Mean Squared Error (MSE)—and three $A\beta$-related measures. Specifically, we compute Spearman correlations between predicted and ground-truth mean SUVR values in the cerebral cortex and hippocampus, referred to as Cerebral Amyloid Burden Correlation (CABC) and Hippocampal Amyloid Burden Correlation (HABC), respectively. These regions are selected due to their known association with A$\beta$ accumulation~\cite{hampel2021amyloid}. Lastly, we evaluate binary $A\beta$-positivity classification using Balanced Accuracy (BA) to address class imbalance. To account for possible systematic biases in each method, predicted A$\beta$-positivity is determined by applying a data-driven threshold to the predicted mean cortical SUVR. This threshold is selected on the internal validation set to maximize BA and is held fixed during testing. We provide the threshold values for each method, along with details on the statistical tests performed in our experiments, in the Supplementary Material.

\subsection{Ablation Study}\label{sec:ablation-study}
In this section, we present an ablation study to assess (i) the effect of the LAS hyperparameter $m$ on model performance, and (ii) the contribution of each individual component within the CoCoLIT framework.

\subsubsection{LAS Hyperparameter Analysis}\label{sec:las-study} 
We evaluate the effect of the LAS hyperparameter $m$ on model performance, with results presented in Table~\ref{tab:ablation-study}-A. Image-based metrics consistently improve as $m$ increases. While the rate of improvement slows beyond $m=8$, the overall trend indicates that larger $m$ values enhance structural fidelity. Similarly, A$\beta$-related metrics show substantial gains over the baseline ($m=1$), with both correlation measures (CABC, HABC) and A$\beta$-positivity classification (BA) reaching their highest values at $m=64$. The results also suggest that the estimated burden increasingly aligns with the ground-truth as $m$ grows. Based on these findings, we select $m=64$ as the optimal configuration for all subsequent experiments.

\subsubsection{Evaluating Individual Components} 
We perform an ablation study to evaluate the contribution of key components in our framework, focusing on: (i) the use of LAS at inference time; (ii) a variant of our proposed WISL with constant weight \((\lambda_t = 1 \ \forall t \in [0, T])\), referred to as ISL; and (iii) the proposed time-step-dependent WISL loss. Comparing the results of ISL and WISL allows us to assess the impact of varying \(\lambda_t\) over time, as defined in Section~\ref{sec:proposed}. We define the ``Base'' model as CoCoLIT without LAS (\(m = 1\)) and the ControlNet trained without either ISL or WISL. We then independently assess the contribution of each component by progressively adding them. Results are summarized in Table~\ref{tab:ablation-study}-B. Adding ISL during training leads to consistent improvements in both image-based and A$\beta$-related metrics, suggesting that ControlNet and the decoder benefit from image-space guidance during training. Introducing the time-step-dependent weighting (WISL) yields further gains, particularly in A$\beta$ correlation metrics (Base + WISL vs. Base + ISL). These improvements become even more pronounced when LAS is used, with WISL outperforming ISL across all metrics (Base + LAS + WISL vs. Base + LAS + ISL). We hypothesize that ISL, lacking temporal weighting, may prematurely enforce fine detail generation early in the denoising process, potentially disrupting the learned trajectory. In contrast, WISL aligns supervision with the progressive nature of denoising, thereby preserving generative stability. Finally, LAS itself contributes positively across all configurations, enhancing both image fidelity and A$\beta$-related performance. Based on these findings, we adopt the full CoCoLIT model with LAS and WISL for all subsequent experiments. 

\begin{figure*}[t]
\centering
\includegraphics[width=\textwidth]{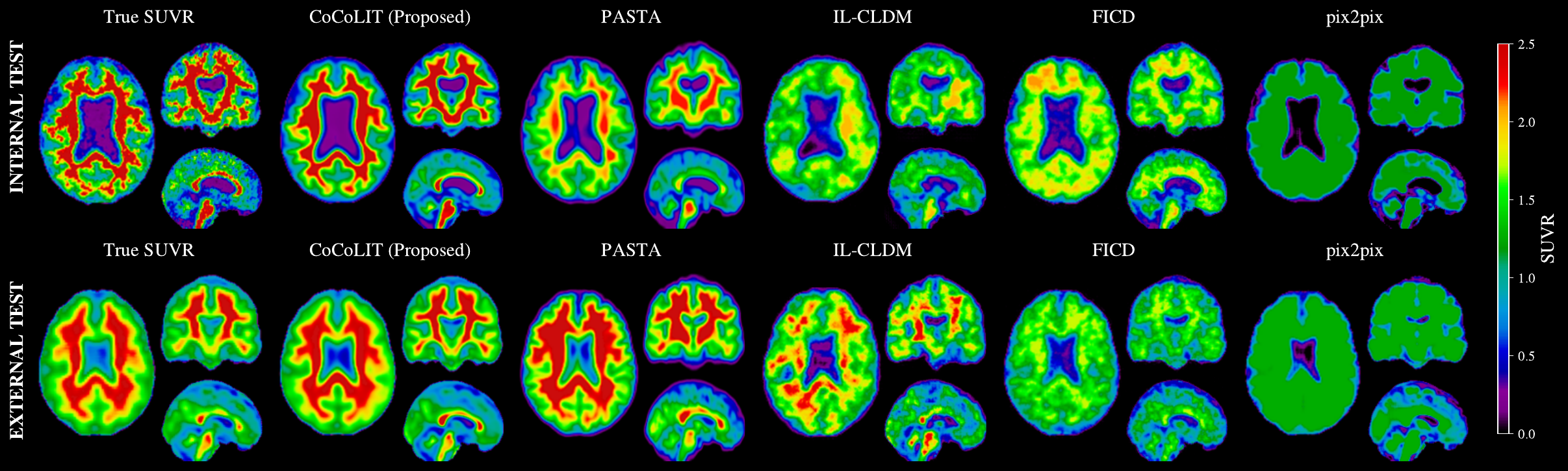}
\caption{Qualitative comparison of SUVR maps predicted from structural MRIs using CoCoLIT and baseline methods on both internal and external test sets. The color bar on the right indicates SUVR values ranging from 0.0 to 2.5.}
\label{fig:qualitative-comparison}
\end{figure*}

\begin{table*}[t!]
    \centering
    \begin{small}
    \begin{tabular}{@{}l rrr rrr@{}} 
    \toprule
    \multicolumn{1}{@{}l}{\textsc{\textbf{Method}}} & \multicolumn{3}{c}{\textsc{\textbf{Image-based Metrics}}} & \multicolumn{3}{c}{\textsc{\textbf{A$\beta$-related Metrics}}} \\
    \cmidrule(lr){2-4} \cmidrule(lr){5-7}
    & \textsc{SSIM}$~\uparrow$ & \textsc{PSNR}$~\uparrow$ & \textsc{MSE}$~\downarrow$ & \textsc{CABC}$~\uparrow$ & \textsc{HABC}$~\uparrow$ & \textsc{BA}$~\uparrow$ \\
    \midrule
    \multicolumn{7}{@{}l}{\scshape\bfseries Internal Test Set} \\ \addlinespace
    pix2pix & $0.693 \pm 0.038$ & $13.968 \pm 1.212$ & $0.0416 \pm 0.0111$ & \underline{$0.178~(p=0.007)$} & $0.363~(p<0.001)$ & \underline{$51.8$\%} \\
    FICD & $0.678 \pm 0.033$ & $12.656 \pm 0.659$ & $0.0549 \pm 0.0084$ & $0.049~(p=0.465)$ & $0.193~(p=0.004)$ & $48.2$\% \\
    IL-CLDM & $0.718 \pm 0.077$ & $18.987 \pm 1.153$ & $0.0131 \pm 0.0038$ & $-0.062~(p=0.357)$ & $0.280~(p<0.001)$ & $46.0$\% \\
    PASTA & \underline{$0.860 \pm 0.042$} & \underline{$21.630 \pm 1.810$} & \underline{$0.0076 \pm 0.0042$} & $-0.006~(p=0.932)$ & \underline{$0.378~(p<0.001)$} & $51.6$\% \\
    CoCoLIT (\textit{Ours}) & $\mathbf{0.896 \pm 0.050}$ & $\mathbf{24.135 \pm 2.820}$ & $\mathbf{0.0050 \pm 0.0047}$ & $\mathbf{0.328}~(p<0.001)$ & $\mathbf{0.522}~(p<0.001)$ & $\mathbf{62.3}$\% \\
    \midrule
    \multicolumn{7}{@{}l}{\scshape\bfseries External Test Set} \\ \addlinespace
    pix2pix & $0.735 \pm 0.035$ & $15.016 \pm 1.198$ & $0.0327 \pm 0.0088$ & \underline{$0.126~(p=0.018)$} & $0.226~(p<0.001)$ & $51.8$\% \\
    FICD & $0.703 \pm 0.028$ & $12.876 \pm 0.629$ & $0.0521 \pm 0.0077$ & $-0.056~(p=0.298)$ & $-0.031~(p=0.558)$ & $49.6$\% \\
    IL-CLDM & $0.744 \pm 0.086$ & $19.918 \pm 1.253$ & $0.0107 \pm 0.0049$ & $0.008~(p=0.879)$ & $0.222~(p<0.001)$ & $50.1$\% \\
    PASTA & \underline{$0.882 \pm 0.028$} & \underline{$22.252 \pm 1.795$} & \underline{$0.0065 \pm 0.0030$} & $0.002~(p=0.967)$ & \underline{$0.235~(p<0.001)$} & \underline{$56.1$\%} \\
    CoCoLIT (\textit{Ours}) & $\mathbf{0.940 \pm 0.010}$ & $\mathbf{26.468 \pm 1.480}$ & $\mathbf{0.0024 \pm 0.0011}$ & $\mathbf{0.801}~(p<0.001)$ & $\mathbf{0.791}~(p<0.001)$ & $\mathbf{79.8}$\% \\
    \bottomrule
    \end{tabular}
    \end{small}
    \caption{Quantitative results from the comparison with baseline methods. Image-based metrics (SSIM, PSNR, MSE) are reported as \textit{mean~$\pm$~std}. A$\beta$ metrics (CABC, HABC, BA) evaluate A$\beta$-burden correlation and classification performance, with p-values for CABC and HABC reported in parentheses. The best result is highlighted in bold, and the second-best is underlined.}
    \label{tab:main-results-updated}
\end{table*}
    
\subsection{Comparison with State-of-the-Art}
In this section, we compare CoCoLIT with several baseline models. We conduct a thorough quantitative evaluation using the metrics described in Section~\ref{sec:evaluation-metrics}, and present qualitative examples of predictions on both internal and external test sets.

\subsubsection{Baselines}
We compare CoCoLIT against existing baseline approaches: PASTA~\cite{li2024pasta}, IL-CLDM~\cite{ou2024image}, FICD~\cite{yu2024functional} and pix2pix~\cite{isola2017image}. All baselines were implemented using their official, publicly available code.

\subsubsection{Quantitative Comparison}
Quantitative results are presented in Table~\ref{tab:main-results-updated}. On the internal test set, CoCoLIT significantly outperforms all baseline methods across both image-based and A$\beta$-related metrics. Notably, while the baselines perform near chance level in A$\beta$-positivity classification, CoCoLIT achieves much better performance, with a BA of 62.3\% (+10.5\% over the second-best method), along with correlations of 0.328 ($p < 0.001$) and 0.522 ($p < 0.001$) for CABC and HABC, respectively. Unexpectedly, on the external test set, all methods exhibit improved image-based metrics and BA. We suspect this is likely due to differences in acquisition protocols and post-processing of PET scans in the A4 study, which result in smoother SUVR signals that are easier to predict. On this dataset, CoCoLIT achieves an SSIM of 0.94, CABC and HABC scores exceeding 0.79, and a BA of 79.8\% (+23.7\% over the second-best method).  All performance improvements are statistically significant both on the internal test set (image-based metrics: $p < 0.001$, BA: $p < 0.05$, except for PASTA [$p < 0.1$]) and on the external test set (image-based metrics: $p < 0.001$, BA: $p < 0.001$). Our results establish CoCoLIT as the new SOTA for MRI-to-PET synthesis and underscore its generalization capabilities, supporting its potential for future clinical translation.

\subsubsection{Qualitative Comparison} 
Figure~\ref{fig:qualitative-comparison} presents visual comparisons of predicted SUVR maps using CoCoLIT and baseline models. Across both the internal and external test sets, CoCoLIT (second column) better approximates the ground-truth A$\beta$ accumulation (first column). In the ground-truth volumes, the smoother SUVR signal observed in the A4 dataset is apparent when compared to ADNI.

\subsection{Empirical Assessment of LAS Theory}
\label{sec:las-linear}
In this section, we empirically evaluate Assumption 1 (see Section~\ref{sec:las}) and compare the performance of the LAS estimator, $\hat{y}$, with the unbiased estimator, $\bar{y}$ (defined in Section~\ref{sec:las-prelim}).

\subsubsection*{Local Linearity of the Decoder (Assumption 1)}
As we show in Section~\ref{sec:las-study}, the LAS bias depends on the decoder's curvature and becomes negligible when the decoder $\mathcal{D}^{(y)}$ is locally linear. We empirically validate this local linearity assumption with two complementary tests. First, we define the experimental setup. For each test subject, we randomly sample five unique pairs of latent vectors, $(z_i^{(y)}, z_j^{(y)})$, from the conditional distribution $p(z^{(y)}|z^{(x)})$. These are decoded to their corresponding outputs, $y_i = \mathcal{D}^{(y)}(z_i^{(y)})$ and $y_j = \mathcal{D}^{(y)}(z_j^{(y)})$. We then construct a linear interpolation path in the latent space using 10 evenly spaced steps $s \in [0, 1]$:
$$z_{\text{interp}}^{(y)}(s) = z_i^{(y)} + s(z_j^{(y)} - z_i^{(y)})$$
The resulting path in the image space is $y_{\text{interp}}(s) = \mathcal{D}^{(y)}(z_{\text{interp}}^{(y)}(s))$. Based on this, we perform two tests.\newline
\textbf{Test 1}: This test assesses whether the distance traveled in the image space increases linearly with the latent interpolation step $s$. We measure this by computing the Pearson Correlation Coefficient (PCC) between the steps $s$ and the corresponding L1 distances from the start point, $d(s) = \|y_{\text{interp}}(s) - y_i\|_1$.\newline
\textbf{Test 2}: This test directly quantifies how much the output path deviates from a perfect straight line. We compare the actual path $y_{\text{interp}}(s)$ to an ideal linear path $\hat{y}_{\text{interp}}(s) = y_i + s(y_j - y_i)$. The deviation is measured as the MSE between these two paths, averaged over all steps $s$.

Across all subjects and latent pairs, we find the mean PCC to be $0.9994 \pm 0.0015$ and the mean MSE to be $0.00045 \pm 0.00057$. Together, these results provide empirical evidence supporting Assumption 1.
\subsubsection{Practical Effectiveness of the LAS estimator}
To evaluate the effectiveness of the LAS estimator, $\hat{y}$, we empirically compare it with the unbiased estimator, $\bar{y}$, which decodes all $N=m=64$ samples before averaging. As shown in Table~\ref{tab:ablation-study}-C, both estimators achieve comparable performance. We therefore conclude that LAS is a practically effective estimator, as its bias does not lead to any meaningful degradation in output quality.

\section{Discussion and Limitations}
In this study, we present CoCoLIT, a novel ControlNet-conditioned latent diffusion framework that outperforms SOTA methods in 3D MRI-to-PET translation. Our method introduces WISL, an image-space supervision loss, and integrates LAS, whose effectiveness is supported by theoretical and empirical results. While our work focuses on MRI-to-PET translation, the CoCoLIT framework is generalizable to a broader range of conditional generative tasks, such as disease progression modeling~\cite{puglisi2025brain}, image quality transfer~\cite{gao2023implicit}, and translation across other imaging modalities~\cite{moschetto2025benchmarking}.

Despite these promising results, some limitations remain. Although the model achieves higher A$\beta$-positivity classification accuracy compared to the SOTA, further improvements may be required to ensure reliable clinical translation. Additionally, although LAS is more efficient than the unbiased estimator at inference time, drawing $m$ samples can still incur computational costs without GPU-parallelization.

In conclusion, future work could explore evaluating CoCoLIT on a broader spectrum of image-to-image tasks, further advancing its synthesis capabilities. Moreover, by leveraging the framework's flexible conditioning mechanism, the integration of clinically relevant covariates may enhance both the predictive power and clinical utility of the model.

\bibliography{aaai2026}

\appendix

\section{Full Derivation of the LAS Estimator Bias}

In this supplement, we provide a detailed, step-by-step derivation for the bias of the LAS estimator, as presented in the main text.
\subsection{Preliminaries and Definitions}
Let $z^{(y)}$ be a $d$-dimensional random latent variable drawn from the conditional distribution $p(z^{(y)}|z^{(x)})$. We define its conditional mean and covariance as:
\begin{align*}
    \mu &= \mathbb{E}[z^{(y)}|z^{(x)}] \\
    \Sigma_{z^{(y)}} &= \mathbb{E}\left[(z^{(y)} - \mu)(z^{(y)} - \mu)^T |z^{(x)}\right]
\end{align*}
From now on, we will omit the conditional dependence on $z^{(x)}$ in all expectations for brevity.

Let $\mathcal{D}^{(y)}: \mathbb{R}^d \to \mathbb{R}^D$ be the decoder network, which we assume to be twice continuously differentiable. The target of our estimation is the expected output, $\mathbb{E}[\mathcal{D}^{(y)}(z^{(y)})]$.

The LAS estimator, $\hat{y}$, is defined by decoding the mean of $m$ i.i.d. samples from the latent distribution:
\[
\hat{y} = \mathcal{D}^{(y)}\left(\bar{z}^{(y)}\right),\quad\text{where}\quad\bar{z}^{(y)}=\frac{1}{m}\sum_{j=1}^m z^{(y, j)},
\]
for samples $z^{(y, j)} \sim p(z^{(y)}|z^{(x)})$.
The bias of this estimator is defined as:
\begin{align*}
    \text{Bias}(\hat{y}) &= \mathbb{E}[\hat{y}] - \mathbb{E}[\mathcal{D}^{(y)}(z^{(y)})]\\ &= \mathbb{E}[\mathcal{D}^{(y)}(\bar{z}^{(y)})] - \mathbb{E}[\mathcal{D}^{(y)}(z^{(y)})].
\end{align*}
    
\subsection{Second-Order Approximation of the True Expectation}
We begin by approximating the expectation, $\mathbb{E}[\mathcal{D}^{(y)}(z^{(y)})]$. We use a second-order Taylor expansion of the vector-valued function $\mathcal{D}^{(y)}$ around the mean latent vector $\mu$.
\begin{equation}
\begin{split}
\label{supp:taylor_single}
\mathcal{D}^{(y)}(z^{(y)}) &\approx \mathcal{D}^{(y)}(\mu) + \nabla\mathcal{D}^{(y)}(\mu)(z^{(y)}-\mu) \\
&\quad+ \frac{1}{2}(z^{(y)}-\mu)^T H_{\mathcal{D}^{(y)}}(z^{(y)}-\mu).
\end{split}
\end{equation}
Here, $\nabla\mathcal{D}^{(y)}(\mu)$ is the Jacobian matrix of $\mathcal{D}^{(y)}$ evaluated at $\mu$, and $H_{\mathcal{D}^{(y)}}$ is the Hessian tensor evaluated at $\mu$. The quadratic form involving the Hessian tensor is a notational shorthand representing the collection of quadratic forms for each output component. Specifically, for the $k$-th component $\mathcal{D}^{(y)}_k$, the second-order term is the standard $\frac{1}{2}(z^{(y)}-\mu)^T H_k (z^{(y)}-\mu)$, where $H_k$ is the Hessian matrix of $\mathcal{D}^{(y)}_k$.

Taking the expectation of both sides of Equation~\eqref{supp:taylor_single}:
\begin{equation}
\begin{split}
\label{supp:taylor_exp}
\mathbb{E}[\mathcal{D}^{(y)}(z^{(y)})] &\approx \mathbb{E}[\mathcal{D}^{(y)}(\mu)] + \mathbb{E}[\nabla\mathcal{D}^{(y)}(\mu)(z^{(y)}-\mu)] \\
&\quad+ \frac{1}{2}\mathbb{E}[(z^{(y)}-\mu)^T H_{\mathcal{D}^{(y)}}(z^{(y)}-\mu)].
\end{split}
\end{equation}
We analyze each term on the right-hand side.
\begin{enumerate}
    \item \textbf{Zeroth-order term:} Since $\mu$ is a constant, $\mathbb{E}[\mathcal{D}^{(y)}(\mu)] = \mathcal{D}^{(y)}(\mu)$.
    \item \textbf{First-order term:} By linearity of expectation, and since $\nabla\mathcal{D}^{(y)}(\mu)$ is a constant matrix:
    \[
    \mathbb{E}[\nabla\mathcal{D}^{(y)}(\mu)(z^{(y)}-\mu)] = \nabla\mathcal{D}^{(y)}(\mu) \mathbb{E}[z^{(y)}-\mu].
    \]
    By definition of $\mu$, we have $\mathbb{E}[z^{(y)}-\mu] = \mathbb{E}[z^{(y)}] - \mu = \mu - \mu = 0$. Thus, the first-order term vanishes.
    \item \textbf{Second-order term:} This term represents a vector in $\mathbb{R}^D$. Let's consider its $k$-th component, where $H_k$ is the $d \times d$ Hessian matrix of the $k$-th output component of the decoder, $\mathcal{D}^{(y)}_k$:
    \[
    \frac{1}{2}\mathbb{E}\left[(z^{(y)}-\mu)^T H_k (z^{(y)}-\mu)\right].
    \]
    The term inside the expectation is a scalar quadratic form. We use the property that for a scalar $s$, $s = \text{Tr}(s)$, and the cyclic property of the trace operator ($\text{Tr}(ABC) = \text{Tr}(CAB)$):
    \begin{align*}
        &\mathbb{E}\left[(z^{(y)}-\mu)^T H_k (z^{(y)}-\mu)\right]\\ &= \mathbb{E}\left[\text{Tr}\left((z^{(y)}-\mu)^T H_k (z^{(y)}-\mu)\right)\right] \\
        &= \mathbb{E}\left[\text{Tr}\left(H_k (z^{(y)}-\mu)(z^{(y)}-\mu)^T\right)\right].
    \end{align*}
    Using the linearity of both expectation and the trace operator, we can swap them:
    \begin{align*}
        &\mathbb{E}\left[\text{Tr}\left(H_k (z^{(y)}-\mu)(z^{(y)}-\mu)^T\right)\right]\\ &= \text{Tr}\left(\mathbb{E}\left[H_k (z^{(y)}-\mu)(z^{(y)}-\mu)^T\right]\right) \\
        &= \text{Tr}\left(H_k \mathbb{E}\left[(z^{(y)}-\mu)(z^{(y)}-\mu)^T\right]\right).
    \end{align*}
    The term $\mathbb{E}\left[(z^{(y)}-\mu)(z^{(y)}-\mu)^T\right]$ is precisely the definition of the covariance matrix $\Sigma_{z^{(y)}}$. Therefore, the $k$-th component of the second-order term is $\frac{1}{2}\text{Tr}(H_k \Sigma_{z^{(y)}})$.
\end{enumerate}
To reassemble the vector from its components, we define the compact notation $\text{Tr}(H_{\mathcal{D}^{(y)}}\Sigma_{z^{(y)}})$ as the vector in $\mathbb{R}^D$ whose $k$-th element is $\text{Tr}(H_k \Sigma_{z^{(y)}})$. This leads to the final approximation for the expectation:
\begin{equation}
\label{supp:true_exp_final}
\mathbb{E}[\mathcal{D}^{(y)}(z^{(y)})] \approx \mathcal{D}^{(y)}(\mu) + \frac{1}{2}\text{Tr}(H_{\mathcal{D}^{(y)}}\Sigma_{z^{(y)}}).
\end{equation}

\subsection{Second-Order Approximation of the LAS Expectation}
Next, we apply the same approximation to the LAS estimator's expectation, $\mathbb{E}[\mathcal{D}^{(y)}(\bar{z}^{(y)})]$. The Taylor expansion is still centered at $\mu$, since $\mathbb{E}[\bar{z}^{(y)}] = \mathbb{E}[\frac{1}{m}\sum_j z^{(y,j)}] = \frac{1}{m}\sum_j \mathbb{E}[z^{(y,j)}] = \frac{1}{m}(m\mu) = \mu$.

The derivation proceeds identically to Section A.2, with the random variable $z^{(y)}$ replaced by $\bar{z}^{(y)}$. This yields:
\begin{equation}
\label{supp:las_exp_interim}
\mathbb{E}[\mathcal{D}^{(y)}(\bar{z}^{(y)})] \approx \mathcal{D}^{(y)}(\mu) + \frac{1}{2}\text{Tr}(H_{\mathcal{D}^{(y)}}\text{Cov}(\bar{z}^{(y)})).
\end{equation}
Our task is to find the covariance of the sample mean, $\text{Cov}(\bar{z}^{(y)})$.
\[
\text{Cov}(\bar{z}^{(y)}) = \text{Cov}\left(\frac{1}{m}\sum_{j=1}^m z^{(y, j)}\right).
\]
Assuming the samples $z^{(y, j)}$ are independent and identically distributed (i.i.d.), the covariance of their sum is the sum of their covariances. Combined with the property that $\text{Cov}(aX)=a^2\text{Cov}(X)$ for a scalar $a$, we have:
\begin{align*}
\text{Cov}\left(\frac{1}{m}\sum_{j=1}^m z^{(y, j)}\right) &= \frac{1}{m^2} \text{Cov}\left(\sum_{j=1}^m z^{(y, j)}\right)\\ &= \frac{1}{m^2} \sum_{j=1}^m \text{Cov}(z^{(y, j)}).
\end{align*}
As $\text{Cov}(z^{(y, j)}) = \Sigma_{z^{(y)}}$ for all $j$, this simplifies to:
\[
\text{Cov}(\bar{z}^{(y)}) = \frac{1}{m^2}\sum_{j=1}^m \Sigma_{z^{(y)}} = \frac{1}{m^2}(m\Sigma_{z^{(y)}}) = \frac{1}{m}\Sigma_{z^{(y)}}.
\]
Plugging this result into Equation~\eqref{supp:las_exp_interim}, we get the approximation for the LAS expectation:
\begin{equation}
\label{supp:las_exp_final}
\mathbb{E}[\mathcal{D}^{(y)}(\bar{z}^{(y)})] \approx \mathcal{D}^{(y)}(\mu) + \frac{1}{2m}\text{Tr}(H_{\mathcal{D}^{(y)}}\Sigma_{z^{(y)}}).
\end{equation}

\subsection{Final Bias Calculation}
Finally, we compute the bias by subtracting the approximation of the true expectation (Equation~\ref{supp:true_exp_final}) from the approximation of the LAS expectation (Equation~\ref{supp:las_exp_final}).
\begin{align*}
\text{Bias}(\hat{y}) &= \mathbb{E}[\mathcal{D}^{(y)}(\bar{z}^{(y)})] - \mathbb{E}[\mathcal{D}^{(y)}(z^{(y)})] \\
&\approx  \frac{1}{2m}\text{Tr}(H_{\mathcal{D}^{(y)}}\Sigma_{z^{(y)}}) - \frac{1}{2}\text{Tr}(H_{\mathcal{D}^{(y)}}\Sigma_{z^{(y)}}) \\
&= \left(\frac{1}{m} - 1\right)\frac{1}{2}\text{Tr}(H_{\mathcal{D}^{(y)}}\Sigma_{z^{(y)}}).
\end{align*}
This completes the derivation of the approximate bias of the LAS estimator.

\section{Data Processing Pipelines}
We preprocess all A$\beta$ PET scans using a bespoke adaptation~\cite{shand2023heterogeneity} of the AmyPET pipeline\footnote{\url{https://github.com/AMYPAD/AmyPET}}. For each subject, we correct the T1-weighted MRI for bias-field inhomogeneities using N4ITK~\cite{tustison2010n4itk}, perform skull-stripping with SynthStrip~\cite{hoopes2022synthstrip}, register the scan affinely to MNI space using ANTs~\cite{avants2008symmetric}, and normalize intensities using the WhiteStripe method~\cite{shinohara2014statistical}. We then align the PET image to the corresponding T1w MRI using affine registration with ANTs. Finally, we generate tissue segmentations with SynthSeg~\cite{billot2023synthseg} to perform SUVR standardization and to quantify amyloid burden correlations in hippocampal and cortical regions. 

\subsection*{Data Dimension}
Both the MRI and PET volumes have dimensions $122 \times 146 \times 122$, and their corresponding latent representations have dimensions $4 \times 32 \times 38 \times 32$.

\section{Details on Evaluation Protocol}

\subsection{Statistical Tests}
Due to the high computational cost, we report results from a single training run for each experiment. Improvements in image-based metrics and balanced accuracy are evaluated using a paired t-test and one-sided bootstrapping, respectively.

\subsection{Data-driven A$\beta$ Classification Thresholds}
We detail all the obtained thresholds for A$\beta$ classification in Table~\ref{tab:method_thresholds}.

\begin{table}[h!]
    \centering
    \renewcommand{\arraystretch}{1.2}
    \begin{tabular}{|l|c|}
        \hline
        \textbf{Method} & \textbf{Data-driven Threshold} \\
        \hline
        pix2pix & 1.102793 \\
        FICD & 1.441395 \\
        IL-CLDM & 1.047870 \\
        PASTA & 1.062552 \\
        CoCoLIT & 1.326200 \\
        \hline
    \end{tabular}
    \caption{Data-driven thresholds for each method, maximizing Balanced Accuracy (BA) for A$\beta$-positivity classification on the internal validation set.}
    \label{tab:method_thresholds}
\end{table}

\end{document}